\documentclass[prl,aps,twocolumn,superscriptaddress,showpacs,preprintnumbers,
               amsmath,amssymb,floatfix]{revtex4}

\def\gtwid{\mathrel{\raise.3ex\hbox{$>$\kern-.75em\lower1ex\hbox{$\sim$}}}}
\def\ltwid{\mathrel{\raise.3ex\hbox{$<$\kern-.75em\lower1ex\hbox{$\sim$}}}}

\usepackage{graphicx}
\usepackage{dcolumn}
\usepackage{bm}

\def\gev{GeV/c$^2$}

\begin{document}


\title{A Search for WIMPs with the First Five-Tower Data from CDMS}

\affiliation{Department of Physics, Brown University,
Providence, RI 02912, USA}
\affiliation{Department of Physics, California Institute of Technology, Pasadena, CA 91125, USA}
\affiliation{Department of Physics, Case Western
Reserve University, Cleveland, OH  44106, USA}
\affiliation{Fermi National Accelerator Laboratory,
Batavia, IL 60510, USA}
\affiliation{Lawrence Berkeley National Laboratory,
Berkeley, CA 94720, USA}
\affiliation{Department of Physics, Massachusetts Institute of Technology, Cambridge, MA 02139, USA}
\affiliation{Department of Physics, Queen's University, Kingston, Ont., Canada, K7L 3N6}
\affiliation{Department of Physics, Santa Clara University, Santa
Clara, CA 95053, USA}
\affiliation{Department of Physics, Stanford University,
Stanford, CA 94305, USA}
\affiliation{Department of Physics, Syracuse University, Syracuse, NY 13244}\affiliation{Department of Physics, University of
California, Berkeley, CA 94720, USA}
\affiliation{Department of Physics, University of
California, Santa Barbara, CA 93106, USA}
\affiliation{Departments of Phys. \& Elec. Engr., University of
Colorado Denver, Denver, CO 80217, USA}
\affiliation{Department of Physics, University of Florida, Gainesville, FL 32611, USA}
\affiliation{School of Physics \& Astronomy, University of Minnesota,
Minneapolis, MN 55455, USA}
\affiliation{Physics Institute, University of Z\"urich, Z\"urich, Switzerland}

\author{Z.~Ahmed} \affiliation{Department of Physics, California Institute of Technology, Pasadena, CA 91125, USA}
\author{D.S.~Akerib} \affiliation{Department of Physics, Case Western Reserve University, Cleveland, OH  44106, USA} 
\author{S.~Arrenberg} \affiliation{Physics Institute, University of Z\"urich, Z\"urich, Switzerland} 
\author{M.J.~Attisha} \affiliation{Department of Physics, Brown University, Providence, RI 02912, USA} 
\author{C.N.~Bailey} \affiliation{Department of Physics, Case Western Reserve University, Cleveland, OH  44106, USA} 
\author{L.~Baudis} \affiliation{Physics Institute, University of Z\"urich, Z\"urich, Switzerland} 
\author{D.A.~Bauer} \affiliation{Fermi National Accelerator Laboratory, Batavia, IL 60510, USA} 
\author{J.~Beaty} \affiliation{School of Physics \& Astronomy, University of Minnesota, Minneapolis, MN 55455, USA} 
\author{P.L.~Brink} \affiliation{Department of Physics, Stanford University, Stanford, CA 94305, USA} 
\author{T.~Bruch} \affiliation{Physics Institute, University of Z\"urich, Z\"urich, Switzerland} 
\author{R.~Bunker} \affiliation{Department of Physics, University of California, Santa Barbara, CA 93106, USA} 
\author{S.~Burke} \affiliation{Department of Physics, University of California, Santa Barbara, CA 93106, USA} 
\author{B.~Cabrera} \affiliation{Department of Physics, Stanford University, Stanford, CA 94305, USA} 
\author{D.O.~Caldwell} \affiliation{Department of Physics, University of California, Santa Barbara, CA 93106, USA} 
\author{J.~Cooley} \affiliation{Department of Physics, Stanford University, Stanford, CA 94305, USA} 
\author{P.~Cushman} \affiliation{School of Physics \& Astronomy, University of Minnesota, Minneapolis, MN 55455, USA} 
\author{F.~DeJongh} \affiliation{Fermi National Accelerator Laboratory, Batavia, IL 60510, USA} 
\author{M.R.~Dragowsky} \affiliation{Department of Physics, Case Western Reserve University, Cleveland, OH  44106, USA} 
\author{L.~Duong} \affiliation{School of Physics \& Astronomy, University of Minnesota, Minneapolis, MN 55455, USA} 
\author{J.~Emes} \affiliation{Lawrence Berkeley National Laboratory, Berkeley, CA 94720, USA} 
\author{E.~Figueroa-Feliciano}\affiliation{Department of Physics, Massachusetts Institute of Technology, Cambridge, MA 02139, USA}
\author{J.~Filippini} \affiliation{Department of Physics, University of California, Berkeley, CA 94720, USA} 
\author{M.~Fritts} \affiliation{School of Physics \& Astronomy, University of Minnesota, Minneapolis, MN 55455, USA} 
\author{R.J.~Gaitskell} \affiliation{Department of Physics, Brown University, Providence, RI 02912, USA} 
\author{S.R.~Golwala} \affiliation{Department of Physics, California Institute of Technology, Pasadena, CA 91125, USA} 
\author{D.R.~Grant} \affiliation{Department of Physics, Case Western Reserve University, Cleveland, OH  44106, USA} 
\author{J.~Hall} \affiliation{Fermi National Accelerator Laboratory, Batavia, IL 60510, USA} 
\author{R.~Hennings-Yeomans} \affiliation{Department of Physics, Case Western Reserve University, Cleveland, OH  44106, USA} 
\author{S.~Hertel}\affiliation{Department of Physics, Massachusetts Institute of Technology, Cambridge, MA 02139, USA}
\author{D.~Holmgren} \affiliation{Fermi National Accelerator Laboratory, Batavia, IL 60510, USA} 
\author{M.E.~Huber} \affiliation{Departments of Phys. \& Elec. Engr., University of
Colorado Denver, Denver, CO 80217, USA} 
\author{R.~Mahapatra} \affiliation{Department of Physics, University of California, Santa Barbara, CA 93106, USA} 
\author{V.~Mandic} \affiliation{School of Physics \& Astronomy, University of Minnesota, Minneapolis, MN 55455, USA} 
\author{K.A.~McCarthy}\affiliation{Department of Physics, Massachusetts Institute of Technology, Cambridge, MA 02139, USA}
\author{N.~Mirabolfathi} \affiliation{Department of Physics, University of California, Berkeley, CA 94720, USA} 
\author{H.~Nelson} \affiliation{Department of Physics, University of California, Santa Barbara, CA 93106, USA} 
\author{L.~Novak} \affiliation{Department of Physics, Stanford University, Stanford, CA 94305, USA}
\author{R.W.~Ogburn} \affiliation{Department of Physics, Stanford University, Stanford, CA 94305, USA} 
\author{M.~Pyle} \affiliation{Department of Physics, Stanford University, Stanford, CA 94305, USA}
\author{X.~Qiu} \affiliation{School of Physics \& Astronomy, University of Minnesota, Minneapolis, MN 55455, USA} 
\author{E.~Ramberg} \affiliation{Fermi National Accelerator Laboratory, Batavia, IL 60510, USA} 
\author{W.~Rau} \affiliation{Department of Physics, Queen's University, Kingston, Ont., Canada, K7L 3N6}
\author{A.~Reisetter} \affiliation{School of Physics \& Astronomy, University of Minnesota, Minneapolis, MN 55455, USA} 
\author{T.~Saab}\affiliation{Department of Physics, University of Florida, Gainesville, FL 32611, USA}
\author{B.~Sadoulet} \affiliation{Lawrence Berkeley National Laboratory, Berkeley, CA 94720, USA} \affiliation{Department of Physics, University of California, Berkeley, CA 94720, USA}
\author{J.~Sander} \affiliation{Department of Physics, University of California, Santa Barbara, CA 93106, USA} 
\author{R.~Schmitt} \affiliation{Fermi National Accelerator Laboratory, Batavia, IL 60510, USA} 
\author{R.W.~Schnee} \affiliation{Department of Physics, Syracuse University, Syracuse, NY 13244} 
\author{D.N.~Seitz} \affiliation{Department of Physics, University of California, Berkeley, CA 94720, USA} 
\author{B.~Serfass} \affiliation{Department of Physics, University of California, Berkeley, CA 94720, USA} 
\author{A.~Sirois} \affiliation{Department of Physics, Case Western Reserve University, Cleveland, OH  44106, USA} 
\author{K.M.~Sundqvist} \affiliation{Department of Physics, University of California, Berkeley, CA 94720, USA} 
\author{M.~Tarka} \affiliation{Physics Institute, University of Z\"urich, Z\"urich, Switzerland} 
\author{A.~Tomada} \affiliation{Department of Physics, Stanford University, Stanford, CA 94305, USA}
\author{G.~Wang} \affiliation{Department of Physics, California Institute of Technology, Pasadena, CA 91125, USA}
\author{S.~Yellin} \affiliation{Department of Physics, Stanford University, Stanford, CA 94305, USA} \affiliation{Department of Physics, University of California, Santa Barbara, CA 93106, USA}
\author{J.~Yoo} \affiliation{Fermi National Accelerator Laboratory, Batavia, IL 60510, USA} 
\author{B.A.~Young} \affiliation{Department of Physics, Santa Clara University, Santa Clara, CA 95053, USA}

\collaboration{CDMS Collaboration}

\noaffiliation


\begin{abstract}
We report first results from the Cryogenic Dark Matter Search (CDMS~II) experiment
running with its full complement of 30 cryogenic particle detectors at the Soudan 
Underground Laboratory.  This report is based on the analysis of data acquired between October 2006 and July 2007 from 15 Ge detectors (3.75\,kg), giving an effective
exposure of 121.3\,kg-d (averaged over recoil energies 10--100\,keV, weighted for 
a weakly interacting massive particle (WIMP) mass of
60\,\gev). A blind analysis, incorporating improved techniques for event reconstruction and data quality monitoring, resulted in zero observed events.
This analysis sets an upper limit on the WIMP-nucleon spin-independent cross section 
of 6.6$\times10^{-44}$\,cm$^2$ (4.6$\times10^{-44}$\,cm$^2$ when combined with previous CDMS Soudan data) at the 90\% confidence level for a WIMP mass of 60\,\gev.  By providing the best sensitivity for dark matter WIMPs with masses above 42 GeV/c$^2$ , this work significantly restricts the parameter space for some of the favored supersymmetric models.
\end{abstract}


\maketitle

Cosmological  observations~\cite{Cosmol} imply the existence of non-baryonic dark matter that drives structure formation on large scales and dominates galactic and extra-galactic dynamics. Weakly Interacting Massive Particles (WIMPs)~\cite{WIMP}, with masses between a few tens of GeV/c$^2$ and a few TeV/c$^2$,  form a generic class of dark matter candidates, motivated~\cite{BigBPart,Reviews} both by the measured value of the cosmological density and by the need to stabilize the standard model of particle physics at the weak scale. 

WIMPs should be distributed in the halo surrounding the Milky Way and scatter in terrestrial particle detectors~\cite{Goodman:1984dc, Gaitskell:2004gd}.  Their coherent scattering on nuclei should lead to a roughly exponential energy-transfer spectrum with a mean recoil energy in the tens of keV~\cite{Reviews, lewin}.  The event rate is expected to be below 0.1\,event per kilogram of target per day, much smaller than radioactivity rates in most materials. A number of technologies, most based on the identification of nuclear recoils among the electron recoils produced by gammas and betas from radioactivity, are starting to reach this sensitivity level, corresponding to a spin-independent WIMP-nucleon scattering cross-section on the order of 10$^{-43}$\,cm$^2$.
Such ``direct'' searches for WIMP elastic scattering are complementary to ``indirect'' searches for their annihilation products in our galaxy and to searches for new physics (e.g.~supersymmetry or additional dimensions) at particle colliders~\cite{Baltz:2006fm}.

The Cryogenic Dark Matter Search (CDMS~II) operates a total of 19\ Ge (250\,g each) and 11\ Si (100\,g each) solid-state detectors at $\sim$\,40\,mK in the Soudan Underground Laboratory~\cite{zips, prd118}. Each detector is a disk, 7.6\,cm in diameter and 1\,cm thick.  Ionizing radiation produces electrons and holes together with phonons.
These charge carriers are drifted by a small electric field (3\,V/cm) and collected on two concentric electrodes on one flat face. 
Athermal phonons are collected using four circuits of superconducting thin-films, each circuit covering a quadrant on the other flat face.  
The ratio of ionization to phonon recoil energy (``ionization yield'') allows us to discriminate nuclear from electron recoils with a rejection factor of $>$\,$10^4$. Electron recoils within $\sim$10\,$\mu$m of the detector surface suffer from a suppressed ionization signal. The resulting reduction in ionization yield can be sufficient to misclassify a surface electron recoil as a nuclear recoil. Signal timing provides effective discrimination against these events, improving our overall rejection of electron recoils to $>$\,$10^6$.  To reduce the external gamma and neutron backgrounds, the experimental setup~\cite{prd118} also includes passive Pb and polyethylene shielding, which is surrounded by an active scintillator veto to detect cosmogenic muons or showers which could produce events in our detectors.

We report on data from two periods (designated as runs 123 and 124) between October 2006 and July 2007. Improvements made since our previous publications~\cite{akerib06} include deployment of 18 additional detectors, increased exposure, greater cryogenic stability, faster data acquisition, enhanced monitoring and control of data quality, and improved analysis techniques. In this analysis we consider only the Ge detectors for WIMP search.  Of the 19\ Ge detectors, three suffering reduced performance from readout failures and one with relatively poor energy resolution have been left out of the present report.  The remaining 15\ Ge detectors (3.75\,kg) were used for the run 123 analysis. Eight of these detectors were excluded from this WIMP search analysis during the shorter run 124 due to variations in performance between the two runs.   The data from these excluded
Ge detectors and the Si detectors will be analyzed for later publication.

Extensive calibrations with gamma ($^{133}$Ba) and neutron ($^{252}$Cf) sources were used to determine data selection criteria (``cuts'') that define the signal region, monitor detector stability, and characterize detector performance.  Calibration data taken regularly with $^{133}$Ba generated over 28 million electron-recoil events between 10--100 keV, exceeding by a factor of thirty the number of comparable events in the WIMP-search data.  Alternating events from the $^{133}$Ba data were separated into two statistically independent samples to allow unbiased characterization of cut performance. Over 600,000 events were recorded using the $^{252}$Cf source during five separate periods throughout the runs, including more than $10^5$ nuclear recoils used to characterize WIMP acceptance.

Both calibration and WIMP search data were used to study detector stability and identify periods of poor performance. A standard set of one- and two-dimensional event parameter distributions were identified and 
data sets with significant deviations were discarded.  A cut was also developed to remove periods of temporary poor ionization collection, during which bulk-electron recoils may leak into the signal region. Finally, we remove events from the outer part of each detector by a fiducial volume cut based on the partitioning of energy between the two concentric charge electrodes.

The phonon energy scale and the timing of the phonon signals vary slightly depending on the position and energy of each event.  In order to maintain effective surface event rejection we compensate for these variations using an empirical look-up table based on our electron-recoil calibrations. The present analysis incorporates energy dependencies into this look-up table alongside position dependencies for the first time, enabling improved surface event discrimination.

Event reconstruction at large radius remains imperfect due to degeneracies in the phonon position quantities that inform the look-up table. 
A small number of high-radius events suffered from miscalibration due to these degeneracies.  We developed a cut on the position-related phonon quantities 
from calibration data to remove events in problematic regions of the look-up table.

We require that a candidate dark matter event deposit significant energy ($>$\,4$\sigma$ above mean noise) in one and only one detector (``single scatter event''), since WIMPs will not interact more than once in our apparatus. 
All 30 detectors contributed to active vetoing of multiple scatter events at all times.
We further require that a WIMP candidate show no significant activity in the surrounding scintillator veto shield during a $200~\mu s$ window around the trigger.  Candidates must also lie within the 2$\sigma$ region of the nuclear-recoil distribution in ionization yield as determined by neutron calibration. 

\begin{figure}[ht]
\begin{center}
\includegraphics[width=3.25in]{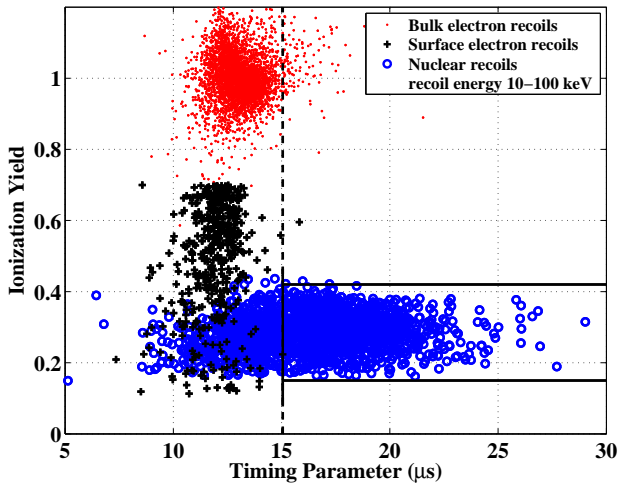} 
\caption{\small  Ionization yield versus timing parameter (see text) for calibration data in one of our Ge detectors. The yield is normalized to unity for typical bulk-electron recoils\,
(dots; from $^{133}$Ba gamma rays). Low-yield $^{133}$Ba events\,(+), attributed to surface electron recoils, have small timing parameter values, allowing discrimination from neutron-induced nuclear recoils from $^{252}$Cf\,($\circ$), which show a wide range of timing parameter values. The vertical dashed line indicates the minimum timing parameter allowed for candidate dark matter events in this detector, and the box shows the approximate signal region, which is in fact weakly energy dependent. (Color online.)}
\label{fig:timing}
\end{center}
\end{figure}

For each event, we measure the risetime of the largest phonon pulse, and also its delay relative to the faster ionization signal.  A cut based on the sum of the risetime and delay provides good rejection of surface electron-recoil events while retaining reasonable acceptance of nuclear recoils. Figure \ref{fig:timing} shows an example of the distribution of this ``timing parameter'' in calibration data (gammas, gamma-induced surface events and neutron-induced nuclear recoils). To effectively remove surface events we require that candidate dark matter events exceed a minimum value for the timing parameter (``timing cut''), determined individually for each detector by setting an allowed maximum passage fraction for surface events in a subset of the $^{133}$Ba calibration data.  We also require that WIMP candidates be consistent with the nuclear-recoil event distribution (i.e. the difference between delay and risetime is less than a 4$\sigma$ deviation from the neutron distribution mean).  The performance of this cut is superior to that of earlier analyses due to improvements to the look-up table.

\begin{figure}[ht]
\begin{center}
\includegraphics[width=3.25in]{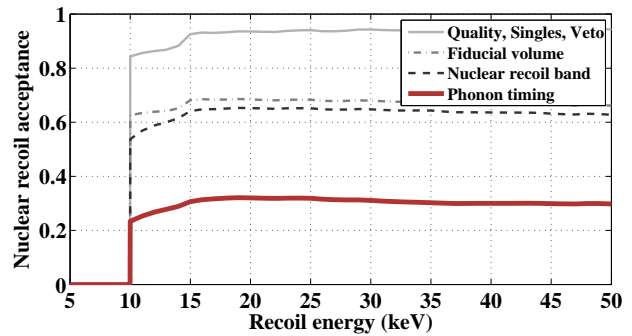}
\caption{\small Nuclear-recoil acceptance efficiency for event-specific cuts (i.e.\ excluding discarded data periods) as a function of recoil energy, averaged over all detectors used in the current analysis, weighted by their individual livetimes.  The four curves represent the cumulative efficiencies at various stages during the analysis, culminating with the final efficiency (bottom) used to generate Figure~\ref{fig:limit}.
}
\label{fig:efficiency}
\end{center}
\end{figure}

The acceptance of our analysis cuts for single-scatter nuclear recoils was measured as a function of energy based on 
both neutron calibration and WIMP search data. Most cuts have very little effect on our acceptance of true nuclear recoils, with the ionization-based fiducial volume and phonon-timing cuts imposing the highest costs in signal acceptance, both measured on neutron calibration data, as shown in Figure~\ref{fig:efficiency}. 
The exposure of this analysis is 397.8\,kg-days before and 121.3\,kg-days after these cuts (averaged over recoil energies 10--100\,keV, weighted for a WIMP mass of 60\,\gev).

To avoid bias, we performed a blind analysis.  An event mask was defined during initial data reduction to remove events in and near the signal region from WIMP search data sets while developing the analysis. This mask was based on primary quantities not subject to refinement during the analysis process, keeping the event selection constant throughout the analysis process described above. After WIMP selection criteria were finalized, the masking was relaxed to cover only the actual signal region to aid in background estimation.

Surface events mainly occur due to radioactive contamination on detector surfaces, or as a result of external gamma ray interactions releasing low-energy electrons from surfaces near the detectors. A correlation analysis between alpha-decay and surface-event rates provides evidence 
that $^{210}$Pb (a daughter-product of $^{222}$Rn) is a major component of our surface event background~\cite{radon}. Surface events generated \emph{in situ} at Soudan, either from calibration with a $^{133}$Ba source or naturally present in the WIMP search data, were studied to understand the surface event leakage into the signal region. We estimate the surface event leakage based on the observed numbers of single- and multiple-scatter events in each detector within and surrounding the $2\sigma$ nuclear-recoil region. The expected background due to surface interactions in this WIMP search analysis is $0.6\pm0.5$ events.

Neutrons induced by radioactive processes or by cosmic-ray muons interacting near the apparatus can generate nuclear-recoil events that cannot be distinguished from possible dark matter interactions on an event-by-event basis. Monte Carlo simulations of the cosmic-ray muons and subsequent neutron production and transport have been conducted with FLUKA~\cite{FLUKA}, MCNPX~\cite{MCNPX} and GEANT4~\cite{GEANT} to estimate this cosmogenic neutron background.  Normalizing the results to the observed veto-coincident multiple-scatter nuclear-recoil rate leads to a conservative
upper limit on this background of $<$0.1 events in our WIMP-search data.

Additional Monte Carlo simulations of neutrons induced by nuclear decay were based on gamma-ray measurements of daughter products of U and Th in the materials of our experimental setup and the assumption of secular equilibrium. The respective background estimate is $<$0.1\,event, dominated by the deduced upper limit of U in the Pb shield. Direct measurements of U in Pb~\cite{EXO_screening} from the same source as the Pb used in our shield suggest a considerably lower contamination.

\begin{figure}[ht]
\begin{center}
\includegraphics[width=3.25in]{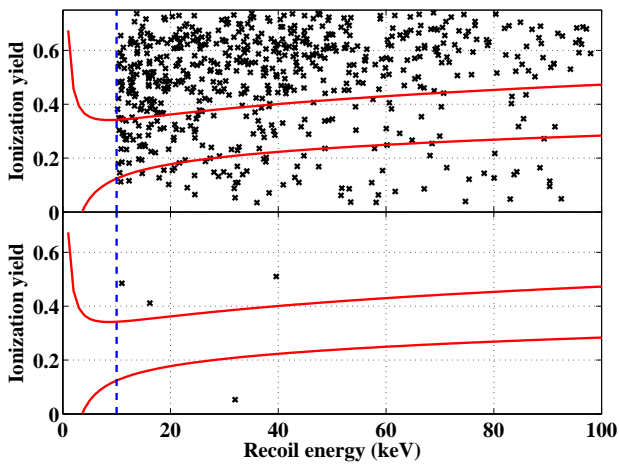}
\caption{\small Top: Ionization yield versus recoil energy in all detectors included in this analysis for events passing all cuts except the ionization yield and timing cuts. The signal region between 10 and 100\,keV recoil energies was defined using neutron calibration data and is indicated by the curved lines. Bulk-electron recoils have yield near unity and are above the vertical scale limits. Bottom: Same, but after applying the timing cut. No events are found within the signal region.}
\label{fig:leakage}
\end{center}
\end{figure}

After all analysis cuts were finalized and leakage estimation schemes selected, we unmasked the WIMP search signal region on February 4, 2008.  No event was observed within the signal region. Figure~\ref{fig:leakage} is a compilation of the low-yield events observed in all detectors used in this analysis. The upper panel shows the ionization yield distribution versus energy for single-scatter events passing all data selection cuts except the timing cut. The four events passing the timing cut shown in the lower panel are outside the 2$\sigma$ nuclear-recoil region.

\begin{figure}[tbp]
\begin{center}
\includegraphics[width=3.25in]{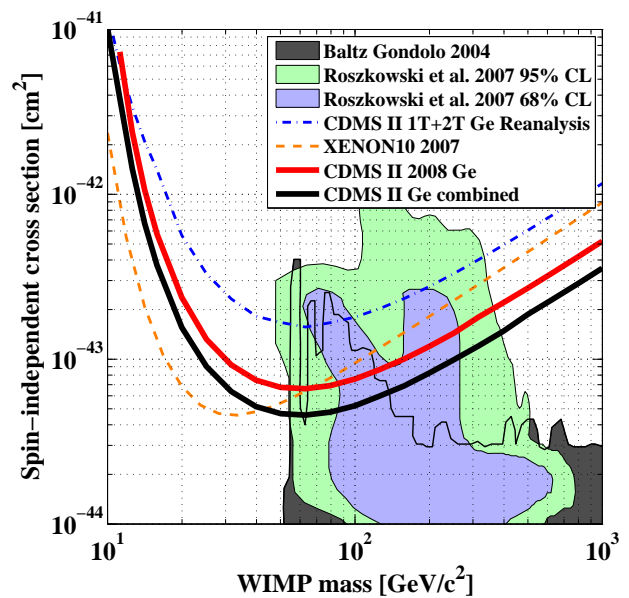}
	\caption{\label{fig:limit}
	\small Spin-independent WIMP-nucleon cross-section upper limits (90\% C.L.) versus WIMP mass.  The upper curve (dash-dot) is the result of a re-analysis~\cite{ogburn} of our previously published data.  The upper solid line is the limit from this work.  The combined CDMS limit (lower solid line) has the same minimum cross-section as XENON10~\cite{Angle:2007uj} (dashed) reports, but has more sensitivity at higher masses.  Parameter ranges expected from supersymmetric models described in~\cite{baltzgondolo} (grey) and~\cite{roszkowskiruiztrotta} are shown (95\% and 68\% confidence levels in green and blue, respectively).  Data courtesy of~\cite{Gaitskell:dmplotter}. }
\end{center}	
\end{figure}

Figure~\ref{fig:limit} shows the Poisson 90\% C.L. upper limit on the spin-independent WIMP-nucleon cross section derived from this data set (upper solid curve), based on standard assumptions about the galactic halo~\cite{lewin}.  The minimum lies at 6.6$\times10^{-44}$\,cm$^2$ for a 60\,\gev ~WIMP.


Our previous data from Soudan~\cite{akerib06, prd118} have been re-analyzed~\cite{ogburn} yielding a slight improvement in sensitivity over our previous publications (upper curve in Figure~\ref{fig:limit}).  A combined limit from all Soudan data (lower solid curve in Figure~\ref{fig:limit}), using Yellin's Optimum Interval method~\cite{Yellin:2002xd} to account for observed events, gives an upper limit of $4.6 \times 10^{-44}$\,cm$^2$ at 90\% C.L. for a WIMP mass of 60\,\gev, a factor of $\sim$3 stricter than our previously published limit.  

We also analyzed our data in terms of spin-dependent WIMP-nucleon interactions. Under the assumption of spin-dependent coupling to neutrons alone and using the Ge form factor given in~\cite{Engel:1991wq}, we find a minimum upper limit of $2.7 \times 10^{-38}$\,cm$^2$ ($1.8 \times 10^{-38}$\,cm$^2$) at 90\% C.L. for this data set (combined Soudan data).

CDMS has maintained high dark matter
discovery potential by limiting expected backgrounds to less than one event in the signal region.  These results from our Soudan measurements set the best WIMP sensitivity for spin-independent WIMP-nucleon interactions over a wide range of WIMP masses. Our new limits cut significantly into previously unexplored regions of the central parameter space predicted by supersymmetry.     

The CDMS collaboration gratefully acknowledges Patrizia Meunier, Daniel Callahan, Pat Castle, Dave Hale, Susanne Kyre, Bruce Lambin and Wayne Johnson for their contributions.  This work is supported in part by the National Science Foundation (Grant Nos.\ AST-9978911, PHY-0542066, PHY-0503729, PHY-0503629,  PHY-0503641, PHY-0504224 and PHY-0705052), by the Department of Energy (Contracts DE-AC03-76SF00098, DE-FG02-91ER40688, DE-FG03-90ER40569, and DE-FG03-91ER40618), by the Swiss National Foundation (SNF Grant No. 20-118119), and by NSERC Canada (Grant SAPIN 341314-07).


%
%
%
%
%
%

\def\journal#1, #2, #3, #4#5#6#7{ 
  #1~{\bf #2}, #3 (#4#5#6#7)} 
\def\apl{\journal Appl.\ Phys.\ Lett., }
\def\apj{\journal Astrophys.\ J., }
\def\apjs{\journal Astrophys.\ J.\ Suppl., }
\def\app{\journal Astropart.\ Phys., }
\def\baas{\journal Bull.\ Am.\ Astron.\ Soc., }
\def\ejpc{\journal Eur.\ J.\ Phys.\ C., }
\def\lnp{\journal Lect.\ Notes\ Phys., }
\def\nature{\journal Nature, }
\def\nc{\journal Nuovo Cimento, }
\def\nima{\journal Nucl.\ Instr.\ Meth.\ A, }
\def\np{\journal Nucl.\ Phys., }
\def\npps{\journal Nucl.\ Phys.\ (Proc.\ Suppl.), }
\def\pl{\journal Phys.\ Lett., }
\def\prep{\journal Phys.\ Rep., }
\def\pr{\journal Phys.\ Rev., }
\def\prc{\journal Phys.\ Rev.\ C, }
\def\prd{\journal Phys.\ Rev.\ D, }
\def\prl{\journal Phys.\ Rev.\ Lett., }
\def\rsi{\journal Rev. Sci. Instr., }
\def\rpp{\journal Rep.\ Prog.\ Phys., }
\def\sjnp{\journal Sov.\ J.\ Nucl.\ Phys., }
\def\solarphys{\journal Solar Phys., }
\def\jetp{\journal J.\ Exp.\ Theor.\ Phys., }
\def\arnps{\journal Annu.\ Rev.\ Nucl.\ Part.\ Sci., }

\end{document}